\documentclass[aps,prd,preprint,noshowpacs,nofootinbib]{revtex4}
\usepackage[dvips]{graphicx} 
\usepackage{amssymb} 
\usepackage{psfrag}
\begin{document}
\title{New inflation, preinflation, and leptogenesis} 
\author{V. N. {\c S}eno$\breve{\textrm{g}}$uz} \email{nefer@udel.edu}
\author{Q. Shafi} \email{shafi@bxclu.bartol.udel.edu} 
\affiliation{Bartol Research Institute, University of Delaware, Newark, DE 19716, USA}
\begin{abstract} 
We present a new inflation model in which the inflationary
scenario including subsequent reheating is determined in part by a $U(1)_R\times
Z_n$ symmetry. A preinflation epoch can be introduced to yield, among other
things, a running spectral index indicated by the WMAP analy\-sis. The inflaton
decay into right handed neutrinos, whose masses can be hierarchical because of
the $Z_n$ symmetry, leads to a reheat temperature $\gtrsim 10^8$ GeV
($\sim10^4$--$10^6$ GeV in some cases), followed
by non-thermal leptogenesis. 
\end{abstract} 
\preprint{BA-04-04} 
\maketitle
\section{Introduction}
Supersymmetric hybrid inflation \cite{Copeland:1994vg} and its extensions (for
reviews see \cite{Lyth:1998xn}) are scenarios that can reconcile the amplitude
of the primordial density perturbations with a GUT scale symmetry breaking,
without any dimensionless parameters that are very small.
An attractive feature of these models is the relative ease with which the
observed baryon asymmetry of the universe can be explained via leptogenesis
\cite{Fukugita:1986hr}, generated by the right handed neutrinos which arise
from the inflaton decay \cite{Lazarides:1997dv}. 

In these models a $U(1)_R$ symmetry plays an essential role in
constraining the form of the superpotential that drives inflation
\cite{Copeland:1994vg}.
A hybrid inflation model with a $U(1)_R\times Z_2$ symmetry, named smooth
hybrid inflation, was considered in \cite{Lazarides:1995vr}, and generalized to
arbitrary $Z_n$ in \cite{Yamaguchi:2004tn}.\footnote{See also \cite{Lyth:1996kt} for related models.} 
In this paper we discuss a related new inflation
model in which the symmetry $U(1)_R\times Z_n$ again
plays an essential role, but unlike smooth hybrid inflation, the inflaton rolls
away rather than towards the origin.  
While the $U(1)_R$ symmetry, as usual, constrains the
superpotential responsible for inflation, the $Z_n$ symmetry also plays an
important role in controlling the reheat process and leptogenesis.

One of our goals in the paper is to realize new inflation in as
natural a manner as possible. This leads us to consider preinflation which is
used to avoid fine tuning the initial conditions. One possible outcome of this
double inflation scenario is a spectral index that can vary significantly with
scale as suggested by the WMAP analysis \cite{Peiris:2003ff}.  We also briefly
discuss the possibility of reconciling early star formation with a running
spectral index \cite{Kawasaki:2003zv,Yamaguchi:2004tn}.

The right handed neutrinos typically acquire
masses through a dimension five term in the superpotential after symmetry
breaking.  For a symmetry breaking scale of order $M_{\rm GUT}$, one would
expect these masses to be of order $M^2_{\rm GUT}/m_P\sim10^{14}$ GeV, where
$m_P=2.4\times10^{18}$ GeV is the (reduced) Planck scale. However, with the
gravitino constraint on the reheat temperature $T_r\lesssim10^{10}$ GeV
\cite{Khlopov:1984pf}, we require the existence of at least one right handed
neutrino of mass $\lesssim10^{12}$ GeV.  The $Z_n$ symmetry is used to ensure
such right handed neutrino masses without invoking small dimensionless
couplings.  

The inflaton decays into right handed neutrinos with masses that
scale as $\mu^p/M_{\rm GUT}^{p-1}$, where $\mu$ is the energy scale of
inflation, and $2<p\le3$. The energy scale is lower in the new
inflation scenario compared to smooth hybrid inflation, which further mitigates
the gravitino problem.  The out of equilibrium decay of the right handed
neutrinos arising from the inflaton satisfactorily explains the observed baryon
asymmetry.\footnote{Leptogenesis in the framework of a similar new inflation model
was considered in \cite{Asaka:1999jb}. Unlike our case, there the inflaton
is a gauge singlet and $Z_n$ is not associated with the neutrino masses.}

The paper is organized as follows. In Sec. II we consider new
inflation and the density fluctuations. In Sec. III we consider the initial condition 
problem where we invoke preinflation as a solution and also as a
way to improve the agreement with the WMAP data. We discuss the right handed
neutrino masses and leptogenesis in Sec. IV, and present our conclusion in Sec.
V.  
\section{New inflation}
We consider the superpotential
\begin{equation} \label{super} 
W_1=S\left(-\mu^2+\frac{(\overline{\Phi}\Phi)^{m}}{M^{2m-2}_{*}}\right)\,, 
\end{equation}
where $\overline{\Phi}(\Phi)$ denote a conjugate pair of superfields
transforming as nontrivial representations of some gauge group $G$, and $S$ is a
gauge singlet superfield. Here $M_*$ is a cut-off scale and $m$ is an integer
 $\ge2$.  Under the $U(1)_R$ symmetry, the superfields transform as
$S\to{\rm e}^{2i\alpha}S$, $\Phi\to\Phi$,
$\overline{\Phi}\to\overline{\Phi}$, and
$W\to{\rm e}^{2i\alpha}W$. Under the discrete symmetry $Z_n$, $\overline{\Phi}$
and $\Phi$ each has unit charge, so that $m=n$ for odd $n$ and $m=n/2$ for
even $n$.  

The vanishing of the F- and D-terms imply that the
SUSY vacua lie at $\langle S\rangle=0$,
$\langle\overline{\Phi}\rangle^*=\langle\Phi\rangle\equiv\pm M$, 
where the symmetry breaking scale $M$ is given by
\begin{equation} \label{def_M}
M=(\mu M^{m-1}_*)^{1/m}\,. 
\end{equation}
With $|\overline{\Phi}|=|\Phi|$ along the D-flat direction of the scalar potential,
we can write the K\"ahler potential as\footnote{We employ 
the same letters for superfields and their scalar components.}
\begin{equation} \label{kahler}
K_1=|S|^2+2\left(|\Phi|^2+\kappa_1\frac{|\Phi|^4}{4m^2_P}+\kappa_2\frac{|S|^2|\Phi|^2}{m^2_P} \right)+
\kappa_3\frac{|S|^4}{4m^2_P}+\ldots\,.
\end{equation}
The scalar potential is given by
\begin{equation}
V={\rm e}^K\left[\left(\frac{\partial^2 K}{\partial z_i\partial z^*_j}\right)^{-1}D_{z_i}W D_{z^*_j}W^*-3|W|^2\right]+V_D\,,
\end{equation}
with
\begin{equation}
D_{z_i}W=\frac{\partial W}{\partial z_i}+\frac{\partial K}{\partial z_i}W\,.
\end{equation}
Using Eqs. (\ref{super}), (\ref{kahler}), and the D-flatness condition, the
scalar potential for $|\Phi|$, $|S|\ll m_P$ is found to be
\begin{equation}
V\simeq\mu^4\left(1-\kappa_3\frac{|S|^2}{m^2_P}+2(1-\kappa_2)\frac{|\Phi|^2}{m^2_P}
-2\frac{|\Phi|^{2m}}{M^{2m}}+\frac{|\Phi|^{4m}}{M^{4m}}\right)\,.
\end{equation}
In contrast to smooth hybrid inflation, where $\kappa_3$ is taken to be
small and positive, for new inflation we take $\kappa_3<-1/3$. The $S$
field acquires a positive mass squared larger than $H^2$, where the Hubble
parameter $H\simeq\mu^2/\sqrt{3}m_P$ during inflation, and therefore rapidly settles to zero.

\begin{figure}[htb] 
\includegraphics[angle=0, width=12cm]{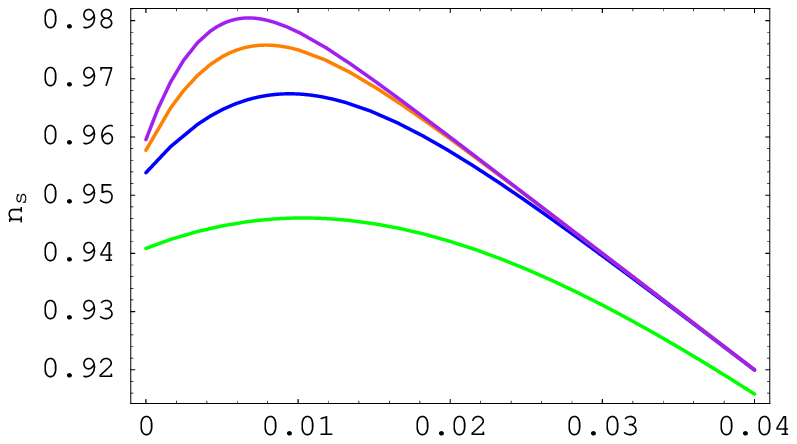} 
\vspace{-0.9cm} 
\begin{center}
{\large \qquad $\beta$} 
\end{center}
 \vspace{-0.8cm} 
\caption{\sf The spectral index $n_s$ vs. $\beta$, for $m=2$ (green), $m=3$ (blue), $m=4$ (orange) and $m=5$ (purple). } \label{fig`1}
\end{figure}

Hereafter we set the symmetry breaking scale $M=M_{\rm
GUT}\simeq2\times10^{16}$ GeV, and use units such that $2M\equiv1$. Defining
$\beta\equiv\kappa_2-1\ge0$ and the canonically normalized real fields
 $\phi\equiv2\rm{Re}\Phi$, $\sigma\equiv\sqrt{2}\rm{Re}S$,
the inflaton potential near the origin is given by
\begin{equation} \label{pot}
V\simeq\mu^4\left(1-\frac{\beta}{2}\frac{{\phi}^2}{m^2_P}-2\phi^{2m}\right)\,.
\end{equation}
This is very similar to the potential of the new inflation model
discussed in \cite{Izawa:1997dv}. The slow roll parameters are given by
\begin{equation} \label{slow}
\epsilon\simeq\frac{1}{2}\left(\beta\frac{\phi}{m_P}+4m m_P \phi^{2m-1}\right)^2\,,
\quad \eta\simeq-\left(\beta+4m(2m-1)m_P^2\phi^{2m-2}\right)\,.
\end{equation}
Since $\epsilon$ is negligible, inflation ends when $|\eta|\simeq1$, at
\begin{equation}
\phi\simeq\frac{1-\beta}{(4m(2m-1)m^2_P)^{1/(2m-2)}}\equiv\phi_f\,.
\end{equation}
The number of $e$-folds after the comoving scale $l$ has crossed the horizon is given by
\begin{equation} \label{e-fold}
N_l=\frac{1}{m^2_P}\int^{\phi_l}_{\phi_f}\frac{V\rm{d}\phi}{V'}
\simeq\frac{1}{(2m-2)\beta}\left[\ln\left(\frac{4m(1+\beta)m^2_P}{1-(2m-2)\beta}\cdot
\frac{\beta+4m m^2_P \phi^{2m-2}_l}{\phi^{2m-2}_l}\right)\right]\,,\ (\beta\not=0),
\end{equation}
where $\phi_l$ is the value of the field at the comoving scale $l$. From Eqs.
(\ref{slow}) and (\ref{e-fold}), for $\beta\gg4m m^2_P \phi^{2m-2}_l$, the
spectral index $n_s$ is found to be
\begin{equation}
n_s\simeq1-2\eta\simeq1-2\beta\left[1+\frac{(2m-1)\left(1-(2m-2)
\beta\right)}{1+\beta}\cdot{\rm e}^{-(2m-2)\beta N_l}\right]\,,
\end{equation}
giving $n_s\simeq1-2\beta$ for $\beta\gg1/[(2m-2)N_l]$. For $\beta=0$ we find
$n_s=1-[2(2m-1)/(2m-2)N_l]$. A numerical calculation of the spectral index, for $m=2$ to 5,
is given in Fig. \ref{fig`1}. Here the values of $n_s$ correspond to the comoving
scale $l_0=2\pi/k_0$, with $k_0\equiv0.002$ Mpc$^{-1}$. The number of
$e$-folds $N_0$ corresponding to this scale is 50--55 depending on the energy
scale $\mu$.\footnote{$N_0\simeq54+(1/3)\ln(T_r/10^{9}\ \rm{GeV})+(2/3)\ln(\mu/10^{14}\ \rm{GeV})$,
where $T_r$ is the reheat temperature.} The running of the spectral index is
negligible, with ${\rm d}n_s/{\rm d}\ln k\lesssim10^{-3}$.

Note that the WMAP analysis suggests a running spectral index, with
$n_s\lesssim0.95$ and ${\rm d}n_s/{\rm d}\ln k\lesssim10^{-3}$ disfavored at
the $2\sigma$ level \cite{Peiris:2003ff}. Other analyses relax this bound to
$n_s\lesssim0.92$ \cite{Seljak:2003jg}, which still requires
$\beta\lesssim0.04$.

\begin{figure}[htb] 
\includegraphics[angle=0, width=12cm]{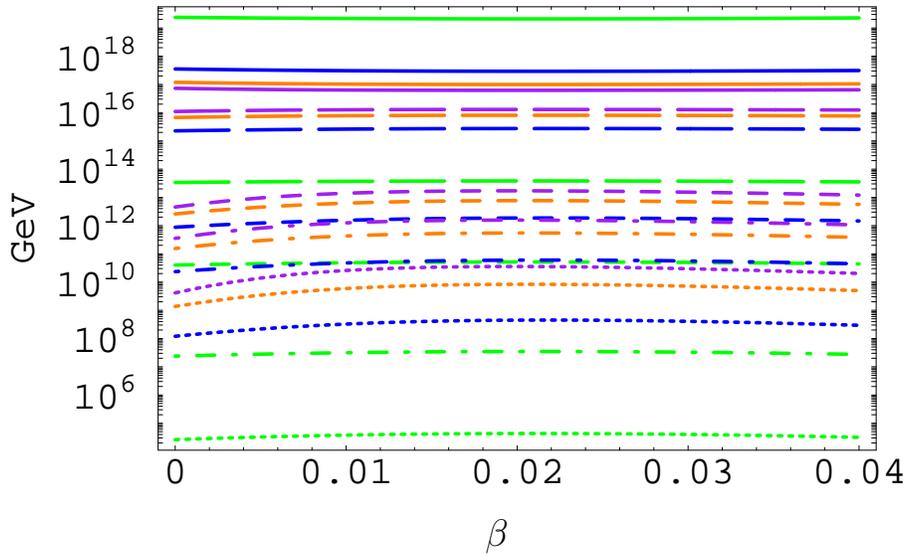} 
\vspace{-0.9cm} 
\begin{center}
{\large \qquad $\beta$} 
\end{center}
 \vspace{-0.8cm} 
\caption{\sf Top to bottom: $M_*$ (solid), $M_3$ (long-dashed), $M_{\theta}$ (dashed), $M_2$ (dot-dashed), and $T_r$ (dotted); for $m=2$ (green), $m=3$ (blue), $m=4$ (orange) and $m=5$ (purple).} \label{fig`2}
\end{figure}

The amplitude of the curvature perturbation $\mathcal{R}$ is given by
\begin{equation}
\mathcal{R}=\frac{1}{2\sqrt{3}\pi m^3_P}\frac{V^{\frac{3}{2}}}{|V'|}
\simeq\frac{\mu^2}{2\sqrt{3}\pi(\beta m_P \phi+4m m^3_P \phi^{2m-1})}\,.
\end{equation}
Using the WMAP best fit $\mathcal{R}\simeq4.7\times10^{-5}$ at $k_0\equiv0.002$ Mpc$^{-1}$
\cite{Peiris:2003ff}, we obtain $\mu\sim2\times10^{13}$ GeV for $m=2$
and $\mu\sim2\times10^{14}$ GeV for $m=5$.  The cut-off scale from Eq.\,(\ref{def_M})
is $M_*\simeq2\times10^{19}$ GeV ($\simeq7\times10^{16}$ GeV) for $m=2$ ($m=5$)
(see Fig. \ref{fig`2}).  

\section{Initial conditions and preinflation}
\begin{figure}[htb]
\includegraphics[angle=0, width=12cm]{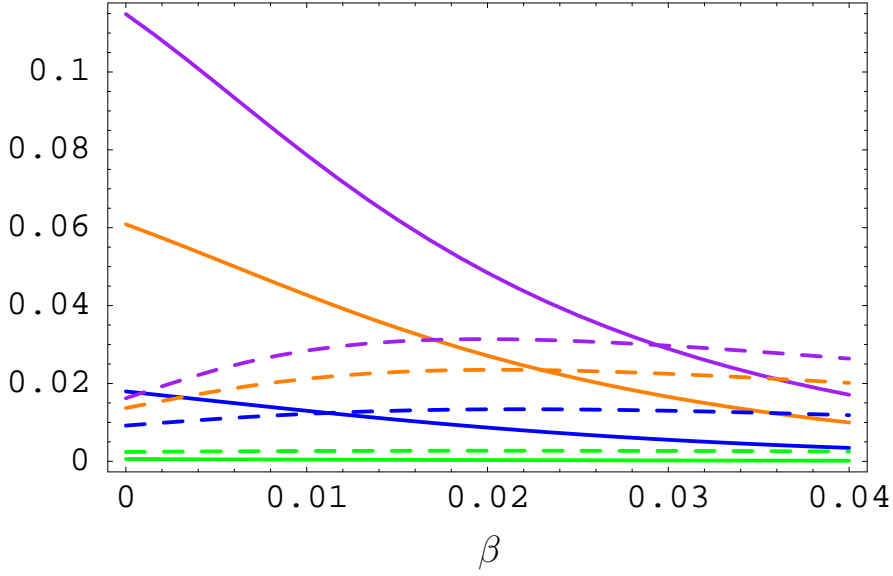} 
\vspace{-0.8cm} 
\begin{center}
{\large \qquad $\beta$} 
\end{center}
 \vspace{-0.8cm} 
\caption{\sf The value of the inflaton field at $k_0=0.002$ Mpc$^{-1}$ in units $2M\equiv1$ (solid curves), and the ratio of the energy scales $\mu/v$ of new inflation and preinflation, where $v\simeq7\times10^{15}$ GeV \cite{Yamaguchi:2004tn} (dashed curves); for $m=2$ (green), $m=3$ (blue), $m=4$ (orange) and $m=5$ (purple).}  \label{fig`3}
\end{figure}

As shown in Fig. \ref{fig`3}, the value of the inflaton field $\phi$ at $k_0$ 
is found to be $\phi_0\sim10^{-4}$ ($10^{-1}$) for $m=2$ ($m=5$),
in units $2M\equiv1$. In other words, the initial value of the inflaton is
close to the local maximum of Eq.\,(\ref{pot}) at $\phi=0$, whereas the true
minimum is at $\phi=1$. Such an initial
condition demands an explanation.

One way to realize the initial condition dynamically 
is through an earlier stage of so-called preinflation \cite{Izawa:1997df}.
Although any suitable preinflation with a high enough energy scale can suppress the initial value of $\phi$, 
to be specific we will assume the superpotential 
to be of the same form as Eq.\,(\ref{super}):
\begin{equation} \label{super2} 
W_2=X\left(-v^2+\frac{(\overline{\Psi}\Psi)^{m}}{M^{2(m-1)}_{S}}\right)\,,
\end{equation}
\begin{equation} \label{kahler2}
K_2=|X|^2+2\left(|\Psi|^2+\alpha_1\frac{|\Psi|^4}{4m^2_P}+\alpha_2\frac{|X|^2|\Psi|^2}{m^2_P} \right)+
\alpha_3\frac{|X|^4}{4m^2_P}+\ldots\,.
\end{equation}
However, unlike $\kappa_3$ in Eq.\,(\ref{kahler}), we take $\alpha_3$ to be positive,
so that smooth hybrid inflation is realized. This particular preinflation stage
has recently been considered in Ref. \cite{Yamaguchi:2004tn}. Note that while
$\alpha_3\lesssim10^{-2}$ is required to obtain the required 50--55 $e$-folds
in a single stage hybrid or smooth hybrid inflation, if the number of $e$-folds
after horizon exit is of order 10, as in the case
of a double inflation model, then $\alpha_3\sim0.1$
is possible.

The inflaton field 
$\phi$ acquires an effective mass during preinflation, given by \cite{Izawa:1997df,Kanazawa:1999ag}
\begin{equation}
m_{\rm eff}=c \frac{v^2}{m_P}=\sqrt{3}c H\,,
\end{equation}
where $v$ is the energy scale of preinflation, and 
$c$ is a parameter which depends on the details of the 
K\"ahler potential. For example, if the latter has a term $2g|\Phi|^2|X|^2$,
the effective mass is equal to $\sqrt{1-g}v^2/m_P$.
The evolution of $\phi$ is given by 
\begin{equation} \label{evolve}
\ddot{\phi}+3H\dot{\phi}+m^2_{\rm eff}\phi=0\,,
\end{equation}
with the solution \cite{Kanazawa:1999ag}
\begin{equation}
\phi\propto {\rm Re}\left[a^{-(3/2)+\sqrt{(9/4)-3c^2}}\right]\,,
\end{equation}
where $a$ is the scale factor of the universe. If we define $\zeta\equiv3/2$ for $c>\sqrt{3}/2$, and
$\zeta\equiv(3/2)-\sqrt{(9/4)-3c^2}$ otherwise, at the end of preinflation $\phi$ takes the value
\begin{equation} \label{phi_e}
\phi\simeq\phi_{\rm min}+(\phi_b-\phi_{\rm min}){\rm e}^{-\zeta N_{\rm pre}}\equiv\phi_e\,.
\end{equation}
Here $\phi_b$ is the value of $\phi$
at the beginning of preinflation, and $N_{\rm pre}$ is the total $e$-fold number of preinflation.
The field $S$ also settles to its minimum the same way. To find these minima we consider
the potential during preinflation, obtained from $W=W_1+W_2$ and 
$K=K_1+K_2$:
\begin{eqnarray} \label{pot_pre}
V&\simeq& v^4\left(1+\frac{|S|^2}{m_P^2}+\frac{2|\Phi|^2}{m^2_P}\right)-
\frac{\mu^2 v^2}{m_P^2}(SX^*+S^* X)\left(1+\frac{(m-1)|\Phi|^{2m}}{M^{2m}}\right)\nonumber\\
&+&\mu^4\left(-\frac{2|\Phi|^{2m}}{M^{2m}}+\frac{|\Phi|^{4m}}{M^{4m}}+\frac{2m^2|S|^2|\Phi|^{4m-2}}{M^{4m}}
\right)\,,
\end{eqnarray}
where we have omitted other higher order terms, and assumed $v>\mu$. 
Identifying the inflaton during preinflation
with $x=\sqrt{2}{\rm Re}X$, Eq.\,(\ref{pot_pre}) is minimized for Im$S$=0. We therefore
can express the potential in terms of the normalized real scalar fields. Eq.\,(\ref{pot_pre}) 
has no minimum for $\phi\ge1$, and for $\phi<1$ it simplifies to
\begin{equation} \label{pot_pre2}
V\simeq v^4\left(1+\frac{\sigma^2}{2m_P^2}+\frac{\phi^2}{2m_P^2}\right)
-\frac{\mu^2 v^2\sigma x}{m_P^2}\,.
\end{equation}
Hence the minima are at $\sigma_{\rm min}\simeq x\mu^2/v^2$, where
$x\sim\langle\Psi\rangle$ at the end of preinflation, and 
$\phi_{\rm min}=0$.\footnote{If $K\supset|\Phi|^2|X|^2+\ldots$, the mass terms
are multiplied with $c$, but this does not affect the minimum of $\phi$
as long as $c$ is positive.}

After preinflation, $W_2$ vanishes and hence the potential has minima at the
origin for both $\phi$ and $\sigma$.  During the matter dominated era between
the two stages of inflation, these fields acquire effective masses $m^2_{\rm
eff}=(3/2)H^2$. Consequently, from Eq.\,(\ref{evolve}) the fields scale as
$a^{-3/4}$ until $H\simeq\mu^2/m_P$.  Since the
energy density scales as $a^{-3}$, the value of $\phi$ at the onset of new inflation is \cite{Kawasaki:1998vx} 
\begin{equation} \label{phi_i}
\phi\simeq\frac{\mu}{v}\phi_e\equiv\phi_i\,.  
\end{equation} 
From Eq.\,(\ref{phi_e}), we see that for $\zeta=3/2$ and $N_{\rm pre}\gtrsim10$,
$\phi_e\cong0$ so that $\phi_i\ll\phi_0$ and new inflation takes
over before the cosmological scales exit the horizon.

Hybrid and smooth hybrid inflation in supergravity provide a natural way to
obtain a blue spectrum at large scales \cite{Linde:1997sj}. Hence, in the light
of the WMAP results, it is of interest to consider the cases where the onset of
new inflation corresponds to cosmological \cite{Kanazawa:1999ag} or
galactic scales, with preinflation responsible for the density fluctuations on
larger scales.  In particular, the latter case can accommodate early star
formation as well as the running spectral index favored by WMAP
\cite{Kawasaki:2003zv,Yamaguchi:2004tn}.  This can be realized in our setup if
we take $c<\sqrt{3}/2$.  As an example, for $c=1/2$ we obtain
$\zeta=(3-\sqrt{6})/2$.  Assuming $\phi_b\simeq m_P$, $N_{\rm pre}\simeq20$ for
the central WMAP values $n_s\simeq1.13$ and ${\rm d}n_s/{\rm d}\ln
k\simeq-0.055$ at $k_0$ \cite{Peiris:2003ff}, and Eq.\,(\ref{phi_e}) gives
$\phi_e\simeq0.25$. Taking $\mu/v=1/50$ in Eq.\,(\ref{phi_i}), we find
$\phi_i\simeq0.005$ (see Fig. \ref{fig`3}).  Higher values of $\phi_i$ are
obtained with slightly lower values of $c$ (e.g. $c=1/3$ yields
$\phi_i\simeq0.12$).   

As shown in Ref. \cite{Kanazawa:1999ag}, for $c>\sqrt{3}/2$, the quantum fluctuations of $\phi$
during preinflation reenter the horizon at the beginning of new inflation with
an amplitude
\begin{equation} \label{del_phi}
\delta\phi\simeq\frac{H_{\rm pre}}{3^{1/4}2\pi}\left(\frac{\mu}{v}\right)^2\,,
\end{equation}
which is a factor $3^{1/4}$ smaller than the fluctuations
induced during new inflation. Moreover, the fluctuations
produced during preinflation are suppressed for smaller wavelengths.
Thus, their contribution to the curvature perturbation can be neglected.

For $c<\sqrt{3}/2$, Eq.\,(\ref{del_phi}) is replaced by
\begin{equation}
\delta\phi\simeq\frac{H_{\rm pre}}{2\pi}\left(\frac{\mu}{v}\right)^{1+2\zeta/3}\,,
\end{equation}
and the corresponding ratio of fluctuations (preinflation/new inflation) is
\begin{equation}
r\equiv\left(\frac{v}{\mu}\right)^{1-2\zeta/3}\,.
\end{equation}
As an example, $c=1/2$ and $v/\mu=50$ yields $r\simeq24$.
Such a crest in the curvature perturbation at scales $\lesssim100$ kpc could 
help resolve the apparent discrepancy between the WMAP predictions of
the running spectral index and early star formation \cite{Spergel:2003cb}.

\section{Neutrino masses and leptogenesis}
The mass matrix $M_R$ of right handed neutrinos is generated
by the superpotential coupling
\begin{equation} \label{coupling}
\gamma_{ij}\left(\frac{\overline{\Phi}\,\Phi}{M^2_*}\right)^{s}
\frac{\overline{\Phi}\,\overline{\Phi}}{M_*}\,N_i\,N_j\,.
\end{equation}
Here $i,j$ denote the family indices, $\gamma_{ij}$
are dimensionless coupling constants assumed to be of order unity, and the vevs of
$\overline{\Phi}$, $\Phi$ along their right
handed neutrino components $\overline{\nu}^c_H$, $\nu^c_H$  break the gauge symmetry.
Eq.\,(\ref{coupling}) yields $M_{R}=\gamma_{ij}M_0 \epsilon^{2s}$; with
$M_0\equiv 2M^2/M_*$ and $\epsilon\equiv M/M_*$.
We assume that $\overline{\Phi}$, $\Phi$ each has unit charge under the discrete symmetry
$Z_n$. Denoting the $Z_n$ charge of $N_i$ by $q_i$, $s$ is given by 
$q_i+q_j+2s+2=2m$. Assigning $q_3=m-1$, $q_1=q_2=0$, the neutrino mass matrices are of the form 
\begin{eqnarray} \label{even}
M_{R} \sim M_0\left(\begin{array}{ccc}
                       \epsilon^{2(m-1)} & \epsilon^{2(m-1)} & 0\\
                      \epsilon^{2(m-1)} & \epsilon^{2(m-1)} & 0\\
                      0 & 0 & 1\\
                      \end{array}\right) ,
\end{eqnarray}
for even $m$, and
\begin{eqnarray}
M_{R} \sim M_0\left(\begin{array}{cccc}
                       \epsilon^{2(m-1)} & \epsilon^{2(m-1)} & \epsilon^{(m-1)}\\
                      \epsilon^{2(m-1)} & \epsilon^{2(m-1)} & \epsilon^{(m-1)}\\
                      \epsilon^{(m-1)} & \epsilon^{(m-1)} & 1\\
                      \end{array}\right) ,  
\end{eqnarray}
for odd $m$, with coefficients of order unity. Denoting
the mass eigenvalues as $\nu^c_i$ with masses $M_i$, we have 
$M_{1\,(2)}=\gamma_{1\,(2)}\epsilon^{2m-2}M_0$ and $M_3=\gamma_3 M_0$. Here
$\gamma_i$ are coefficients of order unity and we take $\gamma_1<\gamma_2=\gamma_3=1$.
Using Eq.\,(\ref{def_M}), we find $M_2=2(M^{-m}\mu^{2m-1})^{1/(m-1)}$ and
$M_3=2(M^{m-2}\mu)^{1/(m-1)}$. $M_2$ and $M_3$ are in the range 10$^7$--10$^{12}$ GeV and
 10$^{13}$--10$^{16}$ GeV respectively, for $2\le m\le5$ (see Fig. \ref{fig`2}). 

The terms responsible for $M_R$ also cause the decay of the inflaton. After inflation ends, the system
performs damped oscillations about the SUSY vacuum. The oscillating inflaton field is 
$\theta=(\delta\overline{\nu}^c_H+\delta\nu^c_H)/\sqrt{2}$ (where $\delta\overline{\nu}^c_H$, 
$\delta\nu^c_H$ are the deviations of $\overline{\nu}^c_H$, $\nu^c_H$ from $M$), 
with mass $M_{\theta}=\sqrt{2}m\mu^2/M$ \cite{14}.
Since $M_{1,2}\le M_{\theta}/2<M_3$, It decays into $\nu^c_{1,2}$ via Eq.\,(\ref{coupling}), 
and the decay width is
\begin{equation} \label{decay}
\Gamma=\frac{1+y^2}{8\pi}\left(\frac{M_2}{M}\right)^2 M_{\theta}\,,
\end{equation}
where $y\equiv M_1/M_2$.  The decay width of $\theta$ to sneutrinos is
$(M_2/M_{\theta})^2\,\Gamma\ll\Gamma$, since $M_2\ll M_{\theta}$. 
(Note that $S$ rapidly settles to its minimum
during inflation and so does not oscillate afterward.)  
For the MSSM spectrum the reheat temperature $T_r$, given by
\begin{equation} \label{reheat} T_r=\left(\frac{45}{2\pi^2
g^*}\right)^{1/4}\sqrt{\Gamma\,m_P}\simeq
\frac{\sqrt{1+y^2}}{16}\frac{\sqrt{m_P\,M_{\theta}}}{M} M_2\,,
\end{equation}
ranges from $10^4$ GeV for $m=2$, to $10^{8}$ GeV for $m=3$
and $10^{10}$ GeV for large $m$. Note that $T_r\sim10^{10}$ GeV
may lead to overproduction of gravitinos \cite{Khlopov:1984pf},
although the gravitino constraint on $T_r$ varies, depending
on the SUSY breaking mechanism and the gravitino mass $m_{3/2}.$\footnote{For gravity
mediated SUSY breaking models with unstable gravitinos of mass $m_{3/2}\simeq0.1$--1 TeV, 
$T_r\lesssim10^7$--$10^9$ GeV \cite{Kawasaki:1995af}, while
$T_r\lesssim10^{10}$ for stable gravitinos \cite{Fujii:2003nr}.  In gauge
mediated models $T_r$ is generally constrained more severely, although
$T_r\sim10^9$--$10^{10}$ GeV is possible for $m_{3/2}\simeq5$--100 GeV
\cite{Gherghetta:1998tq}.}

The inflaton decays into the right handed neutrinos $\nu^c_1$ and $\nu^c_2$
with branching ratios ${\rm Br_1}=y^2/(1+y^2)$ and ${\rm Br_2}=1/(1+y^2)$
respectively. The decay of these neutrinos into lepton and electroweak Higgs
superfields creates a lepton asymmetry, which is partially converted into
baryon asymmetry via the electroweak sphaleron effects
\cite{Fukugita:1986hr}.  Using the experimental value of the
baryon to photon ratio $\eta_B\simeq6.1\times10^{-10}$ \cite{first},
the required lepton asymmetry is found to be $|n_L/s|\simeq2.5\times10^{-10}$
\cite{Khlebnikov:1988sr}.

The lepton asymmetry is given by 
\begin{equation} \label{nls}
\frac{n_L}{s}\simeq\sum_{i,j}\frac{3}{2}\frac{T_r}{M_{\theta}}{\rm
Br_i}\epsilon^{(j)}_i\,, \end{equation} 
where $\epsilon^{(j)}_i$ is the
contribution from the decay of each $\nu^c_i$ with $\nu^c_j$ in the loop.
We assume that $M_1\gg T_r$ so that washout effects are negligible. We
also neglect the small contribution from the decay of the sneutrinos.
The $\epsilon^{(j)}_i$ are given by \cite{Luty:1992un}
\begin{eqnarray} \label{epsilon} \epsilon^{(j)}_i =
  \frac{1}{8\pi} \frac{1}{\left(h^{\dagger} h\right)_{ii}} {\rm Im} 
\left[ \{ \left( h^{\dagger} h \right)_{ij} \}^2 \right] f
\left( \frac{M_j^2}{M_i^2} \right)
 \,, \end{eqnarray}
with
\begin{eqnarray} \label{susy_f} f(x) \simeq \frac{\sqrt{x}}{2}\left[
\ln\left(1+\frac{1}{x}\right)+\frac{2}{x-1}\right] \,.
\end{eqnarray}
Assuming hierarchical Dirac masses $M^D_i$, the leptonic mixings 
in the right handed sector are
small, and we obtain \cite{Akhmedov:2003dg}
\begin{equation}
( h^{\dagger} h )_{ij}\sim\frac{M^D_i M^D_j}{v^2}\,,
\end{equation}
giving
\begin{equation}
\epsilon^{(j)}_i\sim\frac{1}{8\pi}f\left( \frac{M_j^2}{M_i^2} \right)\frac{M^D_j\,^2}{v^2}\,,
\end{equation}
where $v=174$ GeV is the electroweak vev. To estimate the lepton asymmetry we
take the Dirac masses to be $M^D_1\sim0.1$ GeV, $M^D_2\sim1$ GeV, and
$M^D_3\sim100$ GeV. Here the assumption is that the
Dirac masses coincide with the up-type quark masses at tree level, with large
radiative corrections to the first two family masses.\footnote{As discussed in
Ref. \cite{Akhmedov:2003dg}, setting the Dirac masses strictly equal to the
up-type quark masses and fitting to the neutrino oscillation parameters
generally yields strongly hierarchical $\nu^c$ masses, with $M_1\sim10^5$ GeV.
The lepton asymmetry in this case is too small by several orders of
magnitude.}
 
\begin{figure}[htb] 
\includegraphics[angle=0, width=12cm]{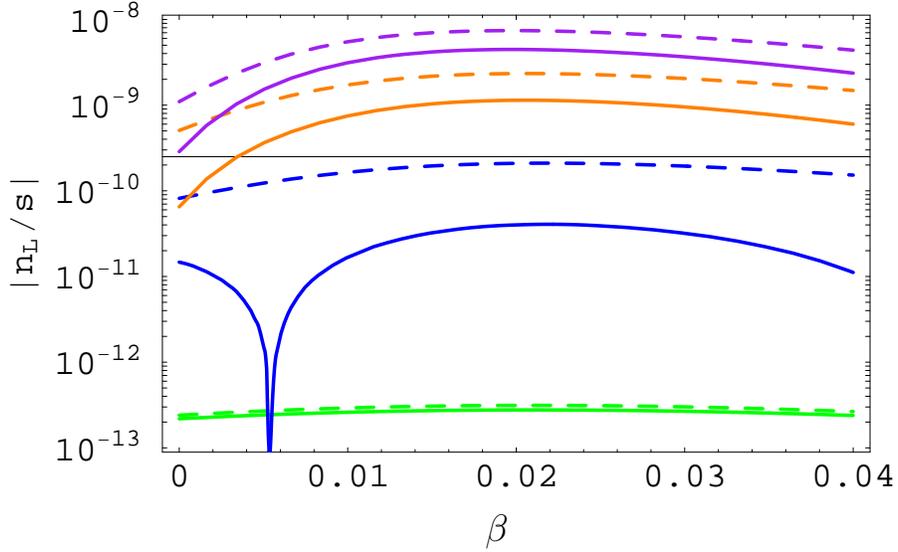} 
\vspace{-0.9cm} 
\begin{center}
{\large \qquad $\beta$} 
\end{center}
 \vspace{-0.8cm} 
\caption{\sf The lepton asymmetry $|n_L/s|$ vs. $\beta$, calculated with $y=0.5$. The horizontal line corresponds to the observed baryon asymmetry. Solid and dashed curves correspond to lower and upper bounds depending on the relative signs of $\epsilon^{(j)}_i$; for $m=2$ (green), $m=3$ (blue), $m=4$ (orange) and $m=5$ (purple).} \label{fig`4}
\end{figure}

The lepton asymmetry calculated with $y=0.5$ is shown in Fig. \ref{fig`4},
where the solid and dashed curves correspond to lower and upper bounds
depending on the relative signs of $\epsilon^{(j)}_i$.  For $m=2$, the
contribution from $\epsilon^{(2)}_1$ dominates and the lepton asymmetry is too
small unless $y\gtrsim.998$.  For $m=3$, the contributions from
$\epsilon^{(2)}_1$ and $\epsilon^{(3)}_2$ are of the same order, and sufficient
lepton asymmetry can be generated with $y\gtrsim0.5$. Finally, for $m>3$ the
contribution from $\epsilon^{(3)}_2$ dominates, and sufficient lepton asymmetry
can be generated as long as $y\gtrsim T_r/M_2\gtrsim10^{-2}$.  The contribution
from $\epsilon^{(3)}_2$ also dominates for $m=2$, provided we do not assign the
$Z_n$ charges as in Eq.\,(\ref{even}) but instead assume $M_i=\gamma_i M_0$
where $\gamma_3\sim1$ and $\gamma_2\sim10^{-4}$. Sufficient lepton
asymmetry can then be generated with $T_r\sim10^6$ GeV and $y\gtrsim
T_r/M_2\gtrsim10^{-3}$.

\section{Conclusion}
We have presented a new inflation model based on a $U(1)_R\times
Z_n$ symmetry. The inflaton is dynamically localized
near the origin prior to new inflation by a stage of
preinflation. With double inflation a running spectral index as well as early
star formation indicated by the WMAP analysis can be accommodated.

The discrete symmetry $Z_n$ enables one to realize hierarchical right handed
neutrino masses without assuming small dimensionless couplings. We show that
sufficient lepton asymmetry to account for the observed baryon asymmetry can be
generated for $m\ge3$, where $m=n$ for odd $n$ and $m=n/2$ for even $n$. The
reheat temperature is found to be $\sim10^8$ GeV for $m=3$ and $\sim10^{10}$
GeV for $m\ge5$. For $m=2$, successful leptogenesis can be realized with a
lower reheat temperatue ($10^4$--$10^6$ GeV), albeit with some small
($10^{-3}$--$10^{-4}$) dimensionless couplings. 
\begin{acknowledgments}
V.N.S. thanks Bumseok Kyae for useful discussions. We thank David H. Lyth
for pointing out a mistake. 
This work is supported by DOE under contract number DE-FG02-91ER40626. 
\end{acknowledgments}

\end{document}